# Empowering Educators in the Age of AI: An Empirical Study on Creating custom GPTs in Qualitative Research Method education

**Qian Huang**, Lee Kuan Yew Centre for Innovative Cities, Singapore University of

Technology and Design, Singapore. Email: qian_huang@sutd.edu.sg

**Thijs Willems**, Lee Kuan Yew Centre for Innovative Cities, Singapore University of

Technology and Design, Singapore, Email: thijs_willems@sutd.edu.sg

## Abstract

As generative AI (Gen-AI) tools become more prevalent in education, there is a growing need to understand how educators, not just students, can actively shape their design and use. This study investigates how two instructors integrated four custom GPT tools into a Master's-level Qualitative Research Methods course for Urban Planning Policy students. Addressing two key gaps: the dominant framing of students as passive AI users, and the limited use of AI in qualitative methods education. The study explores how Gen-AI can support disciplinary learning when aligned with pedagogical intent. Drawing on the Technological Pedagogical Content Knowledge (TPACK) framework and action research methodology, the instructors designed GPTs to scaffold tasks such as research question formulation, interview practice, fieldnote analysis, and design thinking. Thematic analysis of student reflections, AI chat logs, and final assignments revealed that the tools enhanced student reflexivity, improved interview techniques, and supported structured analytic thinking. However, students also expressed concerns about cognitive overload, reduced immersion in data, and the formulaic nature of AI responses. The study offers three key insights: AI can be a powerful scaffold for active learning when paired with human facilitation; custom GPTs can serve as cognitive partners in iterative research practice; and educator-led design is critical to pedagogically meaningful AI integration. This research contributes to emerging scholarship on AI in higher education by demonstrating how empowering educators to design custom tools can promote more reflective, responsible, and collaborative learning with AI.

## Introduction

The rapid development of Generative AI (Gen-AI) technologies has sparked widespread debate about their role in education. Tools such as ChatGPT, Bard, and Claude have demonstrated their capacity to support various aspects of teaching and learning, from content generation to automated feedback (Zhai, 2022). As AI becomes increasingly integrated into classrooms, educators are exploring its pedagogical potential to enhance student engagement and learning outcomes (Holmes et al., 2023). Within the domain of qualitative research methods education, Gen-AI introduces intriguing possibilities for supporting critical thinking, reflexivity, and research skill development.

Much of the existing research on AI in education, however, has focused on **students as passive end-users**, examining how learners interact with AI-based tutoring systems, chatbots, and automated grading systems (Luckin, 2021; Mhlanga, 2023). **Far less attention has been paid to the proactive role educators can play in designing and implementing AI tools** tailored to their specific pedagogical needs. In addition, while AI has been increasingly applied in quantitative disciplines—for example, in data analysis and statistical instruction—**its application in methods education remains underexplored in qualitative research contexts**, where interpretation, reflexivity, and contextual sensitivity are central (Selwyn, 2022).

This study addresses these two key gaps:

1. Most studies on AI in education conceptualize students as (passive) end-users, with limited exploration of educator-driven AI design.
2. The use of AI to support research methods education has been far more common in quantitative disciplines, with relatively little work on its pedagogical role in qualitative research education.

Situated within a Master's-level course on Qualitative Research Methods for Urban Planning Policy students, this study investigates how two instructors designed, implemented, and iteratively refined four custom GPT tools to support student learning. Drawing on the Technological Pedagogical Content Knowledge (TPACK) framework (Mishra & Koehler, 2006), we examine how Gen-AI can be aligned with both disciplinary knowledge and pedagogical strategies to enrich qualitative research education.

This study aims to explore the design, implementation, and pedagogical impact of educator-created Gen-AI tools in qualitative research methods education. Specifically, it addresses the following research questions:

- **RQ1:** How can educators design and implement custom GPT tools to support qualitative research methods education?
- **RQ2:** How do students perceive and engage with these AI-driven tools in their learning process?
- **RQ3:** What insights can be drawn from the development and use of these tools to inform future educator-led AI design and pedagogical innovation in higher education?

By repositioning educators as designers and researchers—not just adopters—this study contributes to ongoing conversations about democratizing AI in education, advancing AI literacy, and supporting discipline-specific pedagogical goals through intentional, reflective use of Gen-AI.

This paper makes three key contributions to the emerging field of AI in education. First, it offers an empirical account of how educators can actively design and integrate Gen-AI tools to support qualitative research pedagogy—an area that remains underexplored compared to quantitative methods instruction. Second, it demonstrates how the TPACK framework can guide the alignment of AI tool design with both disciplinary content and active learning strategies, enabling AI to serve as a scaffold for critical engagement rather than a replacement for human instruction. Third, by analyzing student experiences, challenges, and learning outcomes, the paper provides actionable insights into the affordances and limitations of AI-enhanced education, contributing to broader discussions on AI literacy, human–AI collaboration, and faculty empowerment in the age of educational technology.

**Literature Review**

*AI in Education: Opportunities and Pedagogical Potential*

The integration of Artificial Intelligence (AI) in education has accelerated in recent years, transforming how students interact with knowledge, receive feedback, and develop critical thinking skills (Holmes et al., 2023; Zhai, 2022). AI-powered tools—including intelligent tutoring systems, AI-driven assessments, and generative AI assistants—have demonstrated the ability to personalize learning experiences and enhance student engagement (Luckin, 2021).

One of the most significant developments in this domain is the rise of Generative AI (Gen-AI) models such as ChatGPT, Bard, and Claude. These models can generate text-based responses, summarize readings, and simulate conversations (Kasneci et al., 2023). In higher education, they show promise in supporting complex tasks like academic writing, brainstorming, and research synthesis (Mhlanga, 2023).

These advancements suggest growing pedagogical opportunities for AI in education—particularly when used to support learner autonomy, facilitate reflection, and expand instructional modalities.

*Challenges of AI Use in Educational Contexts*

Despite their promise, Gen-AI tools also raise a number of pedagogical and ethical concerns. One key issue is over-reliance on AI, which may lead to superficial engagement and hinder the development of deep, critical learning (Selwyn, 2022). Additionally, AI-generated content can reflect the biases present in training data, introducing risks related to misinformation, stereotyping, and credibility (Bender et al., 2021).

Another critical issue is the AI literacy gap—many students and educators lack the skills necessary to use AI tools critically and effectively (Tang et al., 2023). Without a clear understanding of how to evaluate, interrogate, and contextualize AI outputs, learners may engage passively rather than thoughtfully with the technology.

These concerns underscore the need for intentional, educator-driven design of AI tools—ensuring they align with pedagogical goals and support deep learning rather than automate surface-level tasks (Holmes et al., 2023).

*The Underexplored Potential of AI in Qualitative Research Education*

While AI has been increasingly integrated into quantitative research education—particularly in areas such as data analysis and statistical instruction—its role in qualitative research education remains limited (Selwyn, 2022). Qualitative research is distinct in its emphasis on interpretation, reflexivity, and context-sensitive analysis (Denzin & Lincoln, 2018). Effective pedagogy in this field requires students to develop competencies in formulating epistemologically grounded questions (Braun & Clarke, 2021), conducting ethically sensitive interviews and observations (Roulston, 2010), and analyzing complex, unstructured data through iterative coding and meaning-making (Saldaña, 2021).

Existant qualitative research pedagogy faces persistent challenges: Students often lack immediate feedback, unlike quantitative tools that yield fast results (Lichtman, 2022). Developing critical reflexivity and understanding positionality are inherently complex and require ongoing guidance (Tracy, 2020).The time-intensive nature of qualitative data collection and analysis makes learning particularly demanding for students and instructors (Creswell & Poth, 2017).

Recent literature points to the potential of AI to support these challenges. For instance, AI can assist students in refining research questions (Kasneci et al., 2023), simulate ethical interview scenarios (Tang et al., 2023), or help categorize field notes (Mhlanga, 2023). However, few studies have examined how educators themselves can design AI tools that are aligned with qualitative pedagogy—this gap is what the present study aims to address.

*Toward Educator-Driven AI Design in Pedagogy*

To ensure AI tools support meaningful learning, educators must be empowered to design them intentionally—tailoring their structure, tone, and interactivity to fit disciplinary and pedagogical goals. Without this proactive design, Gen-AI risks reinforcing biases, perpetuating shallow engagement, and widening literacy gaps (Holmes et al., 2023).

This study builds on that imperative by exploring how custom GPT tools can be created by instructors in the context of qualitative research education. By doing so, it aims to show how AI tools, when embedded in thoughtful pedagogy and domain expertise, can serve as effective, discipline-sensitive scaffolds for learning.

**Theoretical Framework**

To understand how AI can be effectively integrated into qualitative research education, this study employs the Technological Pedagogical Content Knowledge (TPACK) framework (Mishra & Koehler, 2006). TPACK is widely used to analyze technology integration in education, emphasizing the intersection of Content Knowledge (CK, the instructor's expertise in qualitative research methods, including interviewing, fieldwork, and analysis), Pedagogical Knowledge (PK, the instructor's ability to design effective teaching strategies, such as experiential learning, guided reflections, and feedback loops), and Technological Knowledge (TK, the instructor's understanding of AI tools and their application AI tools in teaching.)

TPACK provides a structured lens for analyzing how AI tools are not just technological add-ons but must be deeply embedded in both content and pedagogy. Previous studies using TPACK in AI education (Luckin, 2021; Holmes et al., 2023) have shown, for instance, that effective AI integration requires teachers to balance technology with disciplinary expertise. In other words, AI should enhance, not replace, traditional learning methods, ensuring critical engagement rather than passive AI use. This would require educators to have AI literacy training to develop and customize tools rather than relying solely on pre-built AI systems.

Empirical studies have applied the TPACK framework to explore AI integration in education. For instance, Celik (2023) introduced the **Intelligent-TPACK** framework, extending TPACK to include ethical considerations for AI-based tools. This study developed a scale to measure teachers' knowledge for instructional AI use, emphasizing that technological knowledge must be combined with pedagogical understanding to effectively deploy AI in education. Similarly, Ning et al. (2024) constructed an **AI-TPACK** framework to elucidate the complex interrelations among AI technology, pedagogical methods, and subject-specific content. Their findings suggest that while teachers may exhibit high competence in individual components like Technological Knowledge (TK), they often face challenges in integrating these into comprehensive Technological Pedagogical Content Knowledge (TPACK). These studies underscore the necessity for customized teacher training curricula that enhance integrated knowledge and ethical considerations of AI in teaching.

In the context of this study, the TPACK framework serves as a foundation for designing and implementing custom GPT tools that align with qualitative research education's content and pedagogical requirements. By focusing on the interplay between CK, PK, and TK, educators can create AI-driven tools that not only convey qualitative research methodologies but also engage students through effective pedagogical strategies, ensuring that technology integration enriches the learning experience.

In this study, TPACK is used to analyze how educators 1) designed custom GPTs that align with qualitative research content (CK); 2) implemented AI-enhanced pedagogical strategies to engage students (PK); 3) developed AI prompting techniques and customization skills (TK). By applying TPACK, this study provides a framework for future AI-driven pedagogical innovations, demonstrating how AI can be contextualized within subject-specific education rather than being applied generically.

**Methodology**

*Research Design: Action Research Approach*

This study employs an action research approach (Kemmis & McTaggart, 2000), which is well-suited for examining educational interventions in real-world settings. Action research is a collaborative and iterative methodology that allows educators to develop, implement, and refine teaching practices while engaging in systematic inquiry (McNiff & Whitehead, 2011).

The study follows a plan-act-reflect cycle, wherein: The instructor designed and implemented four custom GPT tools to support qualitative research learning. Students engaged with the tools throughout a one-semester Master's-level course on Qualitative Research Methods for Urban Planning Policy students. Feedback was collected iteratively, allowing adjustments to the AI tools based on student experiences.

By adopting action research, this study ensures that findings are directly applicable to educational practice, emphasizing continuous improvement and reflective teaching (Carr & Kemmis, 1986).

*Participants and Context*

The study was conducted in a Master's-level course on Qualitative Research Methods at a Singapore university. The course was designed for students specializing in Urban Planning Policy, where qualitative research skills are essential for conducting fieldwork, analyzing urban challenges, and engaging with communities.

Participants

- Students (n = 14): Enrolled in the course, from diverse academic and professional backgrounds.
- Instructor (n = 2): The course instructor, who also served as the primary researcher, designed and implemented the custom GPT tools.

To support qualitative research learning, four custom GPTs were developed using OpenAI's GPT-4 plus. Each tool was tailored to specific aspects of the research process, ensuring alignment with course objectives and the TPACK framework (Mishra & Koehler, 2006). Importantly, these tools were not intended to replace in-class activities, but rather to extend and deepen students' learning beyond the classroom environment. The Four Custom GPTs are:

1. QualiQuest Buddy GPT – Helps students refine research questions, explore epistemology, ontology, and positionality, and generate topic ideas.

2. Research Interview Simulator GPT – Simulates interviews, providing adaptive responses, ethical dilemmas, and probing techniques for interview practice.
3. Observation Station GPT – Guides students in analyzing fieldnotes, distinguishing between observation, interpretation, and reflection.
4. DT X Urban Studies GPT – Supports students in applying the Design Thinking Double Diamond Framework to urban challenges.

These GPTs were embedded into post-class activities, offering students opportunities to engage interactively with qualitative research concepts and practice skills outside of formal lecture hours.

*Data Collection Methods*

Data were collected using a multi-method approach, combining qualitative and trace data to gain rich insights into student learning experiences.

| Data Source | Description | Purpose |
| --- | --- | --- |
| Student Reflections (Weekly Journal Entries) | Students recorded their experiences, challenges, and insights after using the GPT tools. | Capture student perceptions and engagement. |
| GPT Conversation Logs | Interaction data from AI chats were anonymized and analyzed. | Examine how students used AI and what types of queries they generated. |
| Final Assignments | Students' final qualitative research reports were analyzed. | Assess whether AI-supported learning translated into improved research skills. |

Using multiple data sources allows for triangulation, enhancing the validity and reliability of findings (Creswell & Poth, 2017).

*Data Analysis: Thematic Analysis Approach*

To analyze the collected data, a thematic analysis approach (Braun & Clarke, 2021) was employed, allowing for a systematic examination of student reflections, GPT interaction logs, and class discussions. This method was chosen for its flexibility in capturing patterns, meanings, and nuanced insights from qualitative data. The analysis followed a structured six-step process. First, the researchers engaged in familiarization with the data by reading and re-reading student responses and AI-generated outputs to develop an initial understanding of emerging themes. Next, initial coding was conducted, where key patterns and recurring insights—such as reflections on AI-enhanced learning, ethical concerns, and challenges in AI usage—were identified. These codes were then grouped into broader themes, including "AI-enhanced reflexivity," "over-reliance on AI," and "ethical AI concerns", ensuring that themes were meaningfully aligned with the research objectives.

Following this, a theme review process was undertaken to ensure coherence, clarity, and non-overlapping categorizations, refining the themes to best represent the dataset. The researchers then defined and named the themes, identifying sub-themes where necessary to provide deeper insights into student experiences with AI-driven learning. Finally, the themes were written up and integrated into the findings section, where direct student quotes and students' AI-generated outputs were used to illustrate key learning patterns and engagement with AI tools. This thematic analysis enabled a nuanced understanding of how custom GPTs influenced student learning, critical thinking, and reflexivity, while also highlighting challenges such as information overload and ethical considerations in AI use.

**Findings**

This section presents the key findings of the study, organized around the three research questions. Each subsection addresses one research question in turn: how educators designed and implemented custom GPT tools (RQ1), how students perceived and engaged with these tools (RQ2), and what broader insights emerged to inform future educator-led AI design and pedagogical innovation (RQ3). Together, these findings offer a comprehensive view of how generative AI can be thoughtfully embedded into qualitative research education through intentional, instructor-driven design.

**RQ1 — Designing and Implementing Custom GPTs through the TPACK Lens**

This study examined how two instructors leveraged the Technological Pedagogical Content Knowledge (TPACK) framework to design and integrate four custom GPT tools into a Master's-level Qualitative Research Methods course. Rather than treating AI as a supplementary technology, the instructors intentionally embedded Gen-AI into their pedagogy, ensuring alignment with disciplinary content, active learning strategies, and technical design. Their process illustrates how AI can be crafted to extend—not replace— human instruction.

Take one of the custom GPTs 'QualiQuest Buddy' as an example. The instructors began by defining the content knowledge (CK) focus: helping students formulate qualitative research questions rooted in ontology, epistemology, and logics of inquiry. From there, they incorporated pedagogical strategies (PK), such as scaffolding feedback, prompting reflexive thinking, and simulating classroom dialogue. Finally, using their technological knowledge (TK), they structured the GPT's interactions—its prompts, responses, and iterative feedback loops—to simulate meaningful dialogue rather than automate answers. This process typified how TPACK principles were applied holistically to design AI tools that meaningfully support qualitative research education.

The instructors' deep knowledge of qualitative research methods shaped the conceptual foundations of each GPT. They identified critical learning areas—such as research design, interview techniques, and fieldwork analysis—where AI could provide structured support. By grounding the tools in core qualitative principles, the educators ensured the GPTs maintained disciplinary fidelity within the TPACK framework. This domain-specific focus allowed AI to enhance methodological understanding rather than dilute it with generic guidance.

To ensure the tools supported student-centered learning, the instructors embedded the GPTs within active pedagogical practices:

- Experiential learning: Students engaged in simulations that mirrored real-world research scenarios.
- Iterative feedback loops: GPTs delivered immediate, adaptive feedback, helping students refine their thinking and techniques.
- Critical engagement: Rather than replacing discussion, GPTs were designed to generate outputs that students and instructors could critique and reflect upon together.

For example, the Research Interview Simulator GPT allowed students to test interview techniques with AI personas and receive real-time suggestions for improvement. Similarly, the Observation Station GPT guided students through analyzing field notes, encouraging them to distinguish between observation, interpretation, and reflection. These tools promoted practice-based learning while retaining the instructor's role in guiding interpretation and critique.

The instructors demonstrated technical adaptability by customizing GPTs without needing programming skills. They used structured prompt engineering and conversation design to shape the AI's behavior, tone, and instructional logic. This no-code customization enabled educators to:Tailor GPT outputs to specific qualitative research processes.Iterate on design based on student feedback.Maintain alignment with the course's epistemological stance.

Unlike traditional educational technologies that require programming, Gen-AI tools like GPT offer low-barrier customization, making pedagogical fluency and content expertise the most important factors for effective implementation. This shift democratizes AI integration in education, allowing instructors—not developers—to shape how AI supports learning.

By aligning the three core domains of TPACK—content, pedagogy, and technology—the instructors created GPTs that were not mere AI tools, but interactive, pedagogically grounded companions in the qualitative research learning journey. This approach demonstrates that when AI design is led by educators with disciplinary insight and pedagogical purpose, Gen-AI can meaningfully extend traditional teaching practices and support deep, reflective learning. *(Try the four custom GPTs the instructors built as shown in Appendix)*

**RQ2 — Student Perceptions and Engagement with AI-Driven Tools**

Students engaged with the custom GPT tools in diverse and meaningful ways throughout the course. Their reflections, assignment data, and AI interaction logs revealed a range of perceptions, benefits, and concerns. The findings below illustrate how students navigated AI-enhanced learning, what they valued, and where they encountered challenges.

**Benefits: How AI Tools Supported Learning**

1) Facilitating Reflexivity and Researcher Self-Awareness

One of the most impactful outcomes was the way AI tools, particularly the Research Interview Simulator GPT, encouraged students to reflect on their own positionality, biases, and ethical responsibilities as qualitative researchers.

> *"In the interview scenario simulations, I realized how my positionality might shape the research process… This self-awareness will guide me in designing interviews that are respectful and open…"* – Student 6

This capacity for critical self-reflection aligns with foundational goals of qualitative research education (Braun & Clarke, 2021), and AI was seen as a non-judgmental space for students to confront complex issues around identity, power, and interpretation.

2) Enabling Quick, Personalized Feedback Loops

Students appreciated the immediacy of AI-generated responses. GPT tools allowed them to receive instant feedback on their interview questions, field notes, and research designs—feedback that was often more timely than what instructors alone could provide.

> *"It allows quick, personalized feedback which reinforces the lessons during class… In the conventional case, we may have already forgotten what we did by the time feedback reaches us."* – Student 3

These real-time exchanges promoted iterative learning, helping students revise their approaches without delay.

3) Enhancing Interviewing Techniques through Simulation

The Research Interview Simulator GPT was especially effective in helping students refine their interviewing strategies. The ability to test and improve questions in a dynamic AI setting gave students valuable practice before conducting real-world interviews.

> *"The simulator made me reflect on my need to be more responsive to participants' answers… and avoid leading questions." – Student 8*

"The suggested improvements helped me modify my questions for subsequent interviews." – *Student 9*

Students highlighted how this AI-driven practice led to greater clarity, depth, and responsiveness in their questioning.

4) Supporting Research Structuring and Analytical Thinking

Several students used the GPTs to organize their research and refine theoretical insights. Tools like DT x Urban Studies GPT and Observation Station GPT helped them apply frameworks, visualize data, and uncover patterns in their notes.

> *"The customized ChatGPT suggested various ways to apply the Design Thinking framework, refine research questions, and visualize collected data." – Student 8*

> *"Running my field notes through the GPT… helped me realize there are actually more interpretations and mini-theories to unpack…" – Student 3*

These interactions showed how AI could serve as a thinking partner, guiding students beyond surface-level analysis.

**II. Challenges: Where AI Fell Short or Required Careful Use**

1) Cognitive Overload from Excessive AI Feedback

While the depth of GPT-generated feedback was appreciated, some students reported feeling overwhelmed by the volume and detail.

> *"It feels a little overwhelming as a lot of information is provided at once sometimes." – Student 9*

This points to the need for moderation and scaffolding in AI feedback to avoid cognitive overload.

2. Reduced Immersion in Qualitative Data

Students expressed concern that AI's efficiency might short-circuit deep, interpretive engagement with data—an essential part of qualitative inquiry.

> *"It may have robbed me of the chance to go through the process of immersing myself in the data..." – Student 3*

This suggests that while AI tools are useful for synthesis, they should not replace manual, reflective data analysis when depth is the learning goal.

3) Dependence on Prompting Skill

Students noted that the quality of AI responses varied significantly depending on how well they framed their prompts.

> *"The quality produced by AI is highly dependent on the quality of prompts... thus the ability to create substantial prompts... is among the skills of the future." – Student 3*

This highlights the need to integrate AI literacy and prompt design into the curriculum to equip students for more effective interactions.

4) Technical Limitations with Access

Several students faced usage limits with the free GPT version, which constrained their engagement with the tools.

> *"With the free version of ChatGPT... I made a second account to continue." – Student 1*

To address this, instructors eventually provided **paid workspace access**, improving equity and tool reliability.

5) Emotional Disconnect and Formulaic Responses

Some students felt that GPT outputs were too mechanical, lacking emotional nuance or the unpredictability of human interactions.

> *"The responses tend to be a bit formulaic... not 'human' enough. The language could tonally match the personas better." – Student 4*

This limitation echoes broader concerns in AI literature about the absence of emotional intelligence in large language models (Bender et al., 2021).

Overall, students engaged deeply with the custom GPT tools, leveraging them to reflect, practice, and improve their research methods. They found value in the immediacy, flexibility, and structured feedback these tools provided. However, they also recognized the need for critical engagement, careful prompting, and human facilitation to avoid over-reliance and ensure meaningful learning. These reflections reinforce the importance of positioning AI not as a replacement for teaching, but as a thoughtfully integrated scaffold that supports, and is shaped by pedagogical intent.

**RQ3 — Insights for Future Educator-Led AI Design and Pedagogical Innovation**

As AI technologies continue to transform higher education, the need for critical, pedagogically grounded engagement with these tools becomes increasingly urgent (Kasneci et al., 2023). This study reveals that while many students and educators currently interact with AI as passive consumers, a more impactful approach involves reimagining both groups as active co-creators of AI-enhanced learning experiences. The findings point to three key insights to guide future educator-led innovation in AI integration.

1. Designing AI as a Scaffold for Active and Reflexive Learning

A central insight from the study is that AI, when intentionally designed, can effectively scaffold active learning and critical reflexivity. Tools such as the *QualiQuest Buddy GPT* and *Research Interview Simulator GPT* gave students structured opportunities to practice research skills, iterate and refine their work,and receive non-judgmental, personalized feedback.

Students reported greater self-awareness of their biases, improved understanding of their positionality, and enhanced confidence in qualitative techniques. These outcomes reflect the transformative potential of AI when it reinforces—not replaces—human-facilitated learning.

However, the findings also underscore the limitations of AI-generated feedback. Students noted that GPT responses sometimes lacked emotional nuance or unpredictability, reinforcing the need for instructor facilitation to support deeper reflection and engagement. Thus, AI's role should be understood as augmenting human teaching—not automating it.

2. Reframing AI as a Cognitive Partner in Human-AI Collaboration

Rather than viewing AI as a static content delivery tool, students and instructors in this study experienced it as a dynamic cognitive partner. When thoughtfully embedded into pedagogy, custom GPTs supported iterative thinking and revision, context-aware dialogue, and personalized exploration of research concepts.

This partnership fostered an interactive learning environment where AI became a thinking companion that responded to students' evolving ideas, questions, and concerns.

Nevertheless, the effectiveness of this human–AI collaboration was contingent on critical human oversight. Students needed guidance to evaluate, reinterpret, and sometimes push back against AI outputs. These findings point toward a model of co-constructed knowledge where learners and educators collaboratively shape learning with AI as an adaptive partner.

The future of AI in higher education lies not in automation, but in dialogic and iterative collaboration between human and machine, facilitated by educator-led design.

3. Empowering Educators to Lead AI Design through Pedagogical Customization

While much research emphasizes student-AI interaction, this study highlights the pivotal role of educators as AI designers. The instructors' deep disciplinary knowledge and pedagogical intent enabled them to create custom GPTs that aligned with specific learning outcomes (e.g., question design, fieldwork analysis), reflected ethical and epistemological values, supported student-centered, inquiry-driven learning.

By leveraging no-code AI customization, instructors shaped the tone, feedback style, and depth of AI interactions. This approach ensured AI functioned not as an external plug-in, but as an extension of the instructors' pedagogical vision.

However, the study also acknowledges a significant barrier: not all faculty possess the confidence or training to design AI tools. To support broader adoption of educator-led AI innovation, institutions must offer accessible, non-technical training in AI tool development, create AI pedagogical labs for interdisciplinary experimentation, and provide sustained investment in infrastructure, resources, and recognition for educator innovation.

Empowering educators in this way repositions them as key agents in shaping how AI is embedded in the future of teaching and learning. This study proposes a new model for AI integration in higher education—one grounded in educator-led design, critical pedagogy, and human–AI collaboration. It shows that AI tools are most effective when developed by educators who align them with disciplinary knowledge, ethical commitments, and active learning strategies. For AI to move beyond automation and toward meaningful educational innovation, institutions must invest in supporting faculty as designers, not just users, of AI. In doing so, AI can become a dynamic partner in building more reflective, responsive, and human-centered learning environments.

**Discussion**

This study contributes to a growing body of research on the pedagogical potential of generative AI (Gen-AI) in higher education by shifting focus from students as AI users to educators as AI designers. While prior work has explored the benefits and risks of AI in education—such as personalization, efficiency, and ethical concerns (Holmes et al., 2023; Bender et al., 2021; Selwyn, 2022)—most studies center on student-AI interaction. This study challenges that paradigm by illustrating how educators can act as **proactive agents** in shaping AI's role through the **design of custom GPT tools** tailored to their content and teaching strategies.

One of the most salient insights is the need to reconceptualize **AI literacy** not merely as a technical competency, but as a **pedagogical and ethical skillset**. Students struggled with crafting effective prompts, critically interrogating AI-generated outputs, and engaging deeply with qualitative data. These challenges underscore earlier warnings in the literature about the risk of over-reliance on AI and superficial engagement (Selwyn, 2022; Tang et al., 2023). This study extends these critiques by emphasizing that AI literacy should also include the ability to assess epistemological assumptions, question interpretive shortcuts, and reflect on the ethics of outsourcing intellectual labor to machines.

From the instructor's perspective, the findings reinforce recent calls for **educator empowerment in AI integration** (Luckin, 2021; Kasneci et al., 2023). The use of the TPACK framework enabled instructors to align AI tool development with disciplinary content (CK), pedagogical intent (PK), and technological adaptability (TK). This approach resonates with Celik's (2023) and Ning et al.'s (2024) emphasis on integrating ethical considerations and pedagogical coherence into AI adoption. By avoiding generic chatbot solutions and creating domain-specific GPTs, instructors ensured that AI served as a **scaffold for student inquiry**, not a shortcut to mechanized output.

Importantly, while students reported enhanced reflexivity, improved research design, and more structured learning, their feedback also revealed critical **limitations**. These included information overload, reduced immersion in data analysis, and an emotional disconnect in AI-generated interactions. These concerns echo broader critiques of Gen-AI's inability to capture **relational depth, contextual subtlety**, and **human unpredictability**—key components of qualitative inquiry (Bender et al., 2021; Tracy, 2020). Thus, while AI can enhance learning, it must be coupled with **human facilitation** to support nuanced interpretation and ethical engagement.

This study also highlights **institutional challenges**. Many educators lack the technical confidence, resources, or support to design and implement custom AI tools. Without targeted faculty development programs, interdisciplinary collaboration, and dedicated infrastructure, educator-led AI innovation risks remaining a niche practice. As such, the findings support broader recommendations for establishing **AI labs**, offering **non-technical training**, and embedding AI literacy into **university policy and pedagogy** (Mhlanga, 2023; Holmes et al., 2023).

**Conclusion**

This study demonstrates the transformative potential of empowering educators to design and implement custom Gen-AI tools. Grounded in the TPACK framework and action research methodology, the research shows that when AI tools are aligned with disciplinary knowledge and pedagogical goals, they can function as powerful thinking partners—supporting reflection, iterative practice, and deeper engagement with qualitative research.

At the same time, the study highlights critical areas for attention: the need for students to develop not just functional, but critical and ethical AI literacies; the importance of human facilitation in guiding reflective use of AI; and the structural support educators require to confidently build, customize, and deploy AI tools.

To move beyond the novelty of AI and toward meaningful integration, higher education must embrace a systemic shift. This includes equipping students to interrogate AI outputs, supporting faculty in developing pedagogically grounded tools, and fostering institutional cultures that prioritize thoughtful, human-centered innovation. Only then can AI serve not just as a technological enhancement, but as a meaningful partner in advancing critical, creative, and collaborative learning.

**Appendix:**

The 4 custom GPTs the instructors/authors built for the course. Scan the QR code to try.

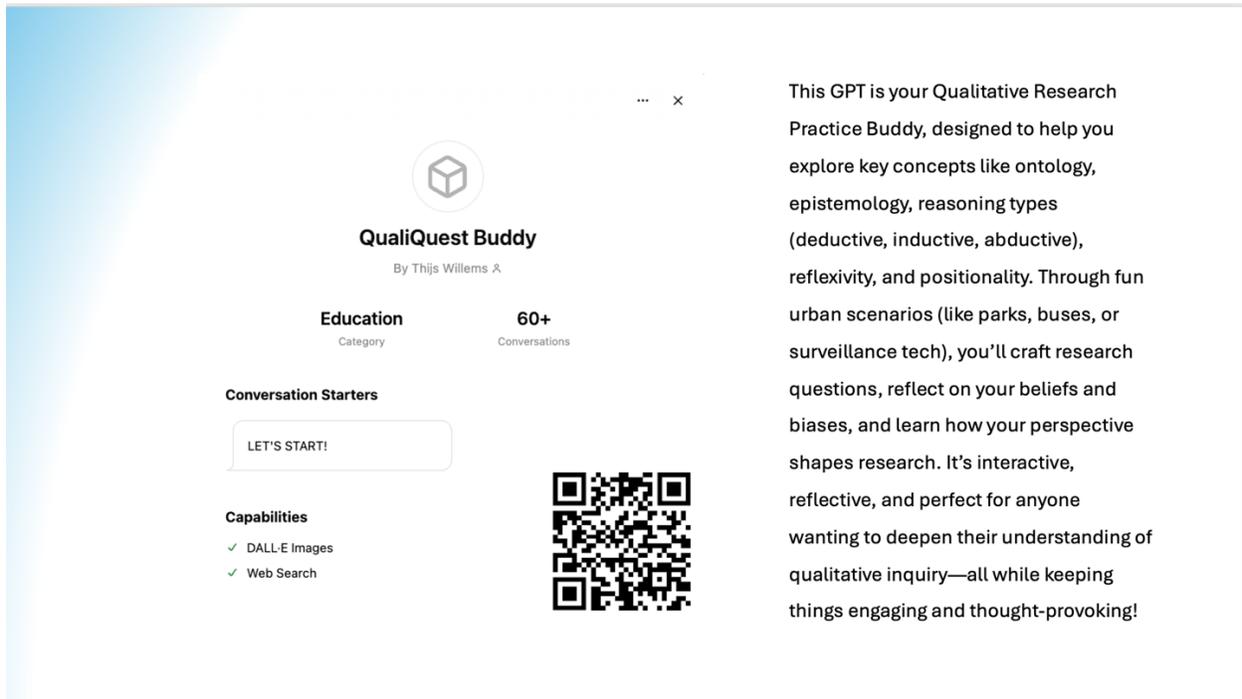

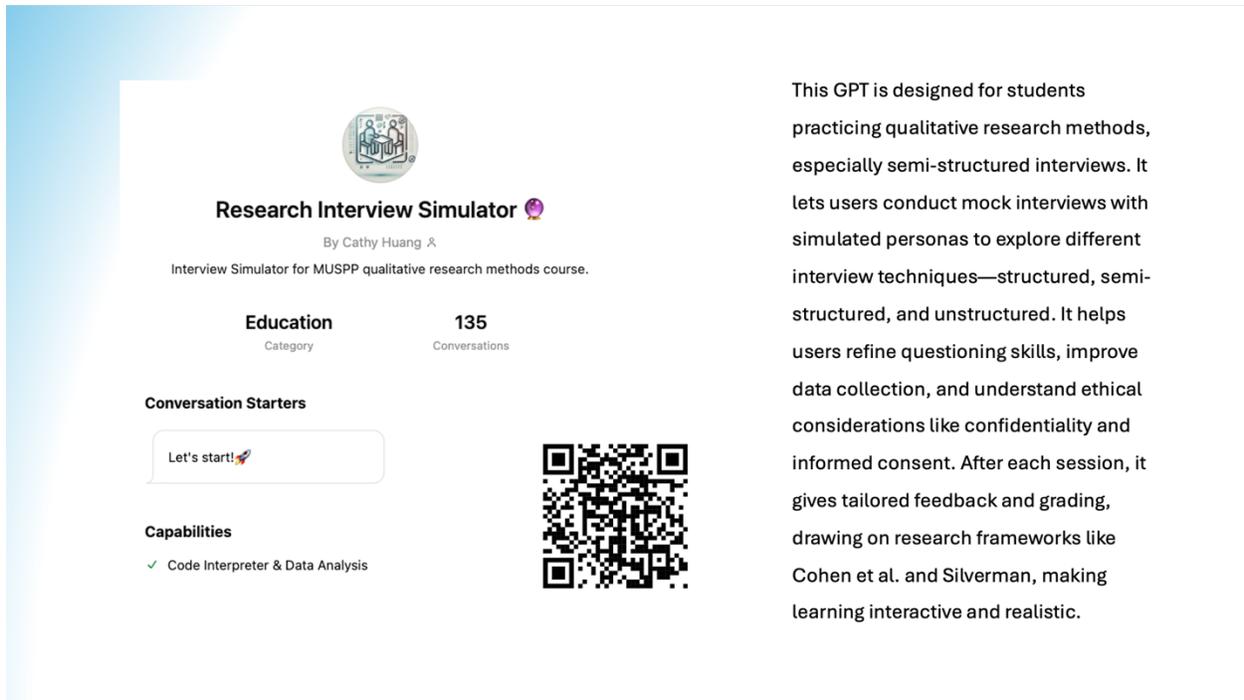

This GPT is designed for students practicing qualitative research methods, especially semi-structured interviews. It lets users conduct mock interviews with simulated personas to explore different interview techniques—structured, semi-structured, and unstructured. It helps users refine questioning skills, improve data collection, and understand ethical considerations like confidentiality and informed consent. After each session, it gives tailored feedback and grading, drawing on research frameworks like Cohen et al. and Silverman, making learning interactive and realistic.

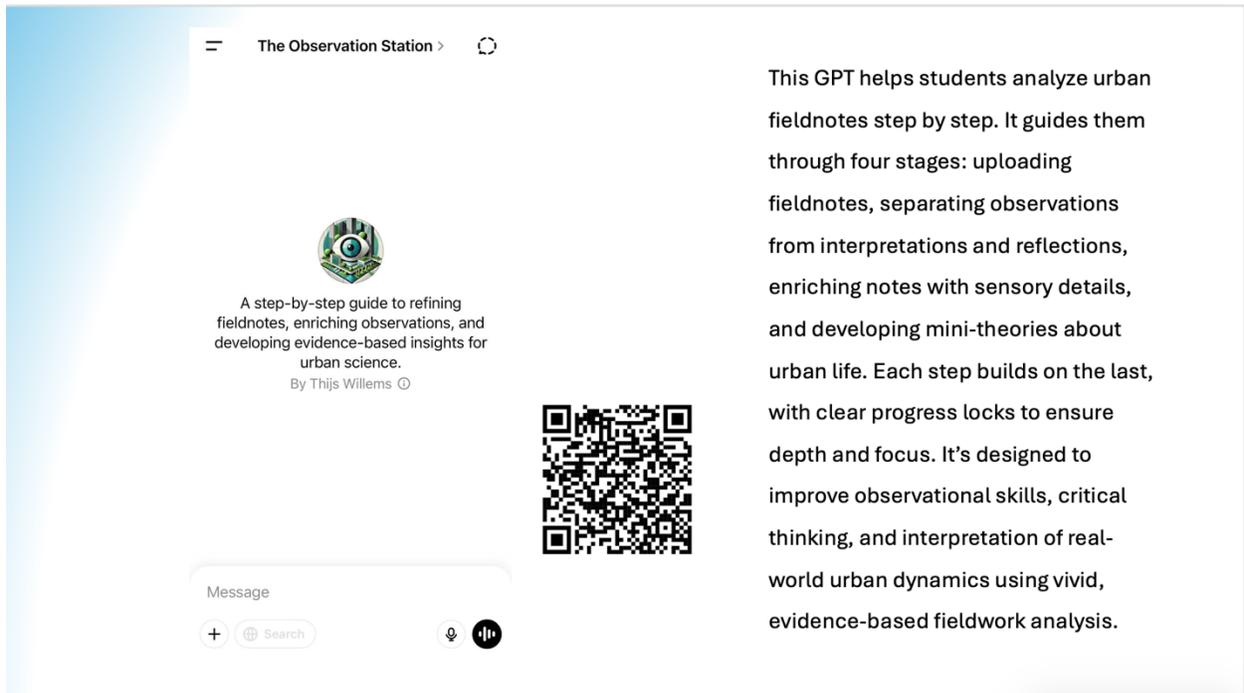

This GPT helps students analyze urban fieldnotes step by step. It guides them through four stages: uploading fieldnotes, separating observations from interpretations and reflections, enriching notes with sensory details, and developing mini-theories about urban life. Each step builds on the last, with clear progress locks to ensure depth and focus. It's designed to improve observational skills, critical thinking, and interpretation of real-world urban dynamics using vivid, evidence-based fieldwork analysis.

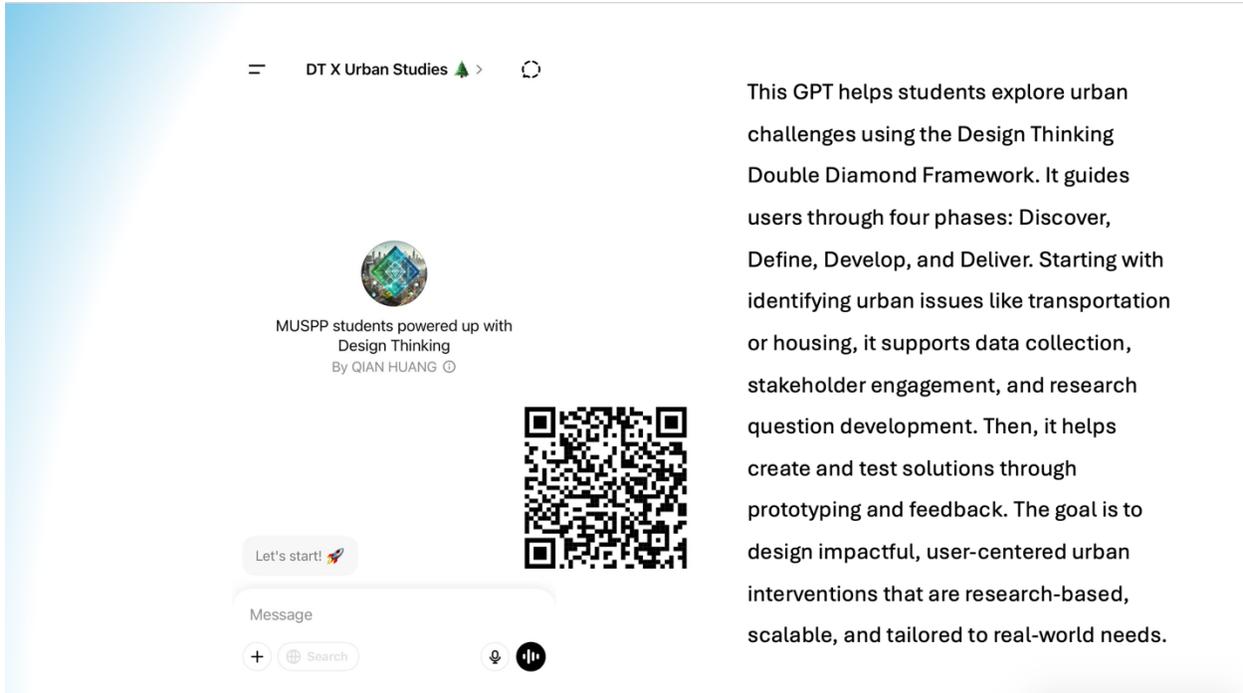